\documentclass[superscriptaddress,iop,apj]{emulateapj}
\usepackage{latexsym}
\usepackage{amssymb}
\usepackage{amsfonts}
\usepackage{threeparttable}                                                               
                            
\usepackage{amsmath}
\usepackage{bm}
\usepackage[normalem]{ulem}
\usepackage{multirow}
\usepackage{subfigure}
\usepackage{color}
\usepackage{calc}
\usepackage{amsmath,amssymb,graphicx}
\usepackage{amssymb,amsmath}
\usepackage{tensor}
\usepackage{bm}
\usepackage{microtype}
\usepackage{booktabs}
\usepackage{times}
\usepackage[varg]{txfonts}
\usepackage{subfigure}
\usepackage{lipsum}
\usepackage{graphicx}
\usepackage{pbox}
\usepackage[colorlinks, citecolor=blue,pdfborder={0 0 0}]{hyperref}

\def\simlt{\mathrel{\rlap{\lower 3pt\hbox{$\sim$}}\raise 2.0pt\hbox{$<$}}}
\def\simgt{\mathrel{\rlap{\lower 3pt\hbox{$\sim$}} \raise 2.0pt\hbox{$>$}}}

\def\gtsima{$\; \buildrel > \over \sim \;$}\def\gtsima{$\; \buildrel > \over
  \sim \;$}

\def\ltsima{$\; \buildrel < \over \sim \;$}
\def\gtrsim{\lower.5ex\hbox{\gtsima}}
\def\lesssim{\lower.5ex\hbox{\ltsima}}
\usepackage{epsfig}
\newcommand{\be}{\begin{equation}}
\newcommand{\ee}{\end{equation}}

\newcommand{\msun}{\,M_{\odot}}

\newcommand{\source}{IGR J17480--2446 }
\newcommand{\q}{\begin{equation}}
\newcommand{\qa}{\begin{eqnarray}}
\newcommand{\qs}{\begin{eqnarray*}}
\newcommand{\nq}{\end{equation}}
\newcommand{\nqa}{\end{eqnarray}}
\newcommand{\nqs}{\end{eqnarray*}}

\begin{document}

\title[Terzan 5] {A Tidal Capture Formation Scenario for the Accreting Pulsar \source in Terzan 5}

\shorttitle{A Young Accreting Pulsar in Terzan 5}

\shortauthors{Patruno \& Mapelli}

\author{A. Patruno \altaffilmark{1,2} \& M. Mapelli \altaffilmark{3}}
\altaffiltext{1}{Leiden Observatory, Leiden University, Neils Bohrweg 2, 2333 CA, Leiden, The 
Netherlands} \altaffiltext{2}{ASTRON, the Netherlands Institute for Radio Astronomy, Postbus 2, 7900 AA, Dwingeloo, the Netherlands}\altaffiltext{3}{INAF-Osservatorio astronomico di Padova, Vicolo dell'Osservatorio 5, I--35122, Padova, Italy}

%\email{patruno@strw.leidenuniv.nl; michela.mapelli@oapd.inaf.it}

\vspace {7cm}

\begin{abstract}
  The low mass X-ray binary (LMXB) \source is an 11 Hz accreting
  pulsar located in the core of the globular cluster Terzan 5. This is
  a mildly recycled accreting pulsar with a peculiar evolutionary
  history since its total age has been suggested to be less than a few
  hundred Myr, despite the very old age of Terzan 5
  (${\sim}12$~Gyr). Solving the origin of this age discrepancy might
  be very valuable because it can reveal why systems like \source are
  so rare in our Galaxy. We have performed numerical simulations
  (dynamical and binary evolution) to constrain the evolutionary
  history of \source. We find that the binary has a high probability
  to be the result of close encounters, with a formation mechanism
  compatible with the tidal capture of the donor star. The result
  reinforces the hypothesis that \source is a binary that started mass
  transfer in an exceptionally recent time.  We also show that
  primordial interacting binaries in the core of Terzan 5 are strongly
  affected by a few hundred close encounters (fly-by) during their
  lifetime. This effect might delay, accelerate or even interrupt the
  Roche lobe overflow (RLOF) phase. Our calculations show that systems
  of this kind can form exclusively in dense environments like
  globular clusters.
\end{abstract}

\section{Introduction}

The 11 Hz accreting pulsar \source is to date the only mildly recycled
accreting pulsar known in our galaxy (see~\citealt{zol17} for an
extra-galactic analogue). This is an accreting neutron star in the
process of being spun-up via channeled accretion and its spin
frequency is smaller than the ${\sim}50$--$100$ Hz required to form an
accreting millisecond X-ray pulsar (AMXP). To date 19 AMXPs have been
identified, with orbital periods ranging from 19 hr down to 40 minutes
and with neutron star spin frequencies between 162 and 599 Hz (see
\citealt{pat12r} for a review and \citealt{str17} for the latest
discovery). In addition, we know in our galaxy three LMXBs and one
intermediate mass X-ray binary (Her X-1) containing slow X-ray pulsars with
spin frequency of 0.1-1 Hz and at least four symbiotic X-ray binaries
with neutron star accretors.
The three slow LMXBs and Her X-1 are different from the 19 AMXPs
because of their strong magnetic dipole moments of the order of
$\mu\simeq10^{30}\rm\,G\,cm^3$ (e.g., \citealt{dai15}), whereas AMXPs
have $\mu\simeq10^{26}\rm\,G\,cm^3$. The mildly recycled pulsar
\source has a dipolar magnetic moment which is constrained to be in
the range $\mu\simeq10^{27}-10^{28}\rm\,G\,cm^3$ and it has an orbital
period of 21.3 hr with a measured mass function giving a minimum donor
mass of 0.4$\msun$ \citep{pap11b, pap12, cav11, pat12b}. It has been
demonstrated that, at least during the outbursts, \source is in a
clear spin-up phase whose magnitude ($\sim10^{-12}\rm\,Hz\,s^{-1}$)
would turn the system into an AMXP in the next few tens of million
years~\citep{pat12b, pap12}.  However, this seems to be at odd with
the very long phases that binaries with parameters similar to \source
spend in Roche lobe contact~\citep{pod02}, which can last for about 1
Gyr or more.~\citet{pat12b} proposed therefore that \source is in an
exceptionally early RLOF phase although the reason on why we are
witnessing this unlikely event remains an open problem.

An interesting property of \source is that it resides in the very
massive and centrally concentrated globular cluster Terzan 5. The
cluster age is constrained to be $12\pm1$ Gyr with a possible second
population of stars, comprising about 40\% of the cluster mass and
formed in a more recent epoch, with an age of $4.5\pm0.5$
Gyr~\citep{fer16,fer09}.  The two populations of stars have different
compositions of $Z=0.01, Y=0.29$ (the old one) and $Z=0.03, Y=0.26$
(the younger one) which is very atypical for globular clusters.
Terzan 5 is also peculiar because it hosts an impressive number of
compact objects, with 37 radio pulsars discovered so far
(\citealt{lyn90, lyn00, ran05, hes06}; see also the on-line
catalog\footnote{\url{http://www.naic.edu/~pfreire/GCpsr.html}}) and at
least 31 X-ray sources.

This large number of compact objects might be connected with the role
of dynamical encounters in the cluster. In particular, the fact that
\source lies within the core radius of Terzan 5 \citep{hei06},
strongly suggests that dynamical encounters might also have played a
significant role in the formation and evolution of this binary, which
could perhaps explain the peculiar evolutionary stage in which it is
currently observed (but see also \citealt{pat12b} and \citealt{tau13}
for a formation scenario that involves accretion induced collapse of a
white dwarf).

\citet{pat12b} provided an analytic study of the three evolutionary
epochs (magnetic-dipole dominated, wind fed and RLOF) that a typical
neutron star LMXB follows during its lifetime and concluded that
\source is in an exceptionally early RLOF phase with a slightly
evolved donor star (a sub-giant) close to the turn off mass of Terzan
5 ($M\sim0.9-1.1\msun$, depending to which of the two stellar
populations does it belong). The total age of the binary was
constrained to be less than a few hundred Myr when considering the
duration of each of the three evolutionary epochs. The evolution of
\source seems to be anomalous in this sense, and its location in the
globular cluster Terzan 5 might help to explain why it is so.
\citet{jia13} in particular, provided a first discussion of this
possibility and suggested that \source\, has been formed by an
exchange interaction in between a binary and an isolated recycled
pulsar (or a binary containing the current pulsar and the current
donor star).

In this paper we present new results that help to further constrain
the history of \source.  We present dynamical and binary evolution
calculations to help determine whether \source is a primordial binary
or whether its accretion history has been strongly affected by the
multiple interactions that occur in the globular cluster core. In
Section~\ref{sec:num} we describe the numerical codes used in our
calculations. In Section~\ref{sec:res} we present our results and we
investigate three possible scenarios for the origin of the system:
exchange interaction (Section~\ref{sec:exc}), tidal capture
(Section~\ref{sec:tid}) and accretion induced collapse
(Section~\ref{sec:aic}). In Section~\ref{sec:rlo} we expand our
results by broadly considering the effect that close encounters have
on the evolution of interacting binaries located in the dense core of
clusters like Terzan 5. In Section~\ref{sec:con} we outline our
conclusions.

\section{Numerical Simulations}\label{sec:num}

We have carried a set of dynamical and binary evolution calculations
to test possible formation scenarios and evolutionary histories of
\source\,. We first begin by describing the dynamical simulations and
then we outline the main features of the binary evolution code used in
this work.

\subsection{Dynamical Simulations}\label{sec:dyn}

Dense concentrated environments like the core of Terzan 5 play a
critical role in determining the evolution of binaries.  We check here
with which probability close encounters affect the evolution of
\source. We try to answer to two questions: 
\begin{itemize}
\item what is the exchange rate for the binaries in the core of Terzan 5 ?
\item What is the rate of close encounters (fly-by) for each binary in the
core of the cluster ? 
\end{itemize}

We use a set of hybrid Monte Carlo and three-body encounter
simulations to perform the integration of 10,000 binaries.  We
adopt the upgraded version of the Binary EVolution (BEV) code by
\citet{sig95} described in \citet{map04, map06} (see also
\citealt{map07, map09}).  The code calculates the dynamics of binaries
under the influence of the host globular cluster potential, 
dynamical friction, and distant and close (i.e., three body)
encounters with other stars.

\subsubsection{Potential, Dynamical Friction and Distant Encounters}

Excluding three-body encounters (whose treatment is different and will
be discussed separately), the binaries are evolved in the globular
cluster center-of-mass frame according to \citep{sig95}
\begin{equation}
\frac{{\rm d}^2{\bf r}}{{\rm d}t^2}=\nabla{}\Psi{}(r)+{\bf a}_{\rm dyf}+{\bf a}_{\rm kick},
\end{equation}
where $\nabla{}\Psi{}(r)$ is the potential gradient due to the mass
interior to $r$, ${\bf a}_{\rm dyf}$ is the dynamical friction
experienced by the binary, and ${\bf a}_{\rm kick}$ is the effective
acceleration due to time-averaged distant encounters with single
stars. To estimate ${\bf a}_{\rm dyf}$ and ${\bf a}_{\rm kick}$, the
diffusion coefficients \citep{bin87} are first calculated in the code.

The potential of the host globular cluster is represented by a time
independent, isotropic multimass King model. We model the population
of Terzan~5 with 10 classes of mass (defined by the mass ranges:
$0.1-0.157$, $0.157-0.20$, $0.20-0.25$, $0.25-0.31$,$0.31-0.39$,
$0.39-0.60$, $0.60-0.70$, $0.70-0.90$, $0.90-1.32$, $1.32-1.57$
M$_\odot{}$). The turn-off (TO) mass is set to be $0.9$ M$_\odot{}$.
To calculate the potential, we input the observed core density
($n_c{\sim}2\times10^6\rm\,pc^{-3}$)  and velocity dispersion
($\sigma{}_c{\sim}15\rm\,km/s$) of Terzan~5~\citep{ori13,pra16}, and we
modify the value of the central adimensional potential, $W_0$ (defined
in \citealt{sig95}), until we reproduce the concentration and the
density profile of Terzan~5.

\subsubsection{Three-body Encounters}

At each step, the probability of a close encounter between the
simulated binary and a single star is evaluated. If the probability is
found to be sufficiently high (on the basis of a random number), the
parameters of the encounter (relative velocity, impact parameter and
orientation angles) are also generated.  The criteria to evaluate the
probability of an encounter, as well as the physical parameters of the
encounter, are described in detail in \citep{sig95}.
Three-body encounters are implemented by means of a 4$^{th}-$order
Runge-Kutta integration scheme with adaptive step size and quality
control, and with the requirement that the angular momentum is conserved
to one part in $10^5$. This guarantees that the energy is conserved to
better than one part in 10$^7$ during the three-body encounter~\citep{sig93}.

\subsubsection{Binary Initial Conditions}

An ensemble of binaries was evolved for a fixed time $t=12$ Gyr, in
the described model of globular cluster. The primary is fixed to be a
neutron star and its initial mass ($m_1$) is set to 1.4
M$_\odot{}$. The mass of the secondary ($m_2$) is drawn at random from
a \citet{sal55} initial mass function, requiring that
$0.1\le{}m_2/{\rm M}_\odot{}\le{}1.2$. The initial semi-major axis
$a_{\rm in}$ is randomly drawn from a uniform $\log{(a)}$
distribution (requiring $7\times{}10^{-3}\le{}a_{\rm in}/{\rm
  AU}\le{}1\times{}10^{2}$). Finally, the initial eccentricity $e_{\rm
  in}$ is randomly drawn from a distribution $P(e)=2\,{}e$
($0\le{}e_{\rm in}<1$).

\subsection{Binary Evolution Calculations}

\begin{figure}
\centering
\rotatebox{-90}{\includegraphics[width=0.7\columnwidth]{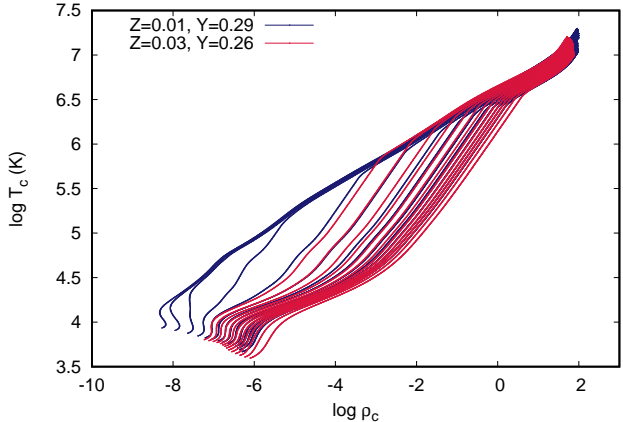}}
\caption{Trajectories of central conditions (x-axis: central density,
  y-axis: central temperature) for pre-main sequence stars evolved
  with {\tt MESA} up to the ZAMS. The different trajectories refer to
  masses from 0.63$\msun$ (bottom curve) to 1.55$\msun$ (top curve) in
  logarithmic steps of 0.03. The red and dark blue lines refer to the
  different compositions used in this work and the end of each line at
  the top right marks the start of the ZAMS. All stellar models used in
  this paper have initial (ZAMS) conditions generated by interpolating the values of the ZAMS models shown. }\label{fig:zam}
\end{figure}

For our binary evolution calculations we use the Modules for
Experiments in Stellar Astrophysics
({\tt MESA};~\citealt{pax11,pax13,pax15}), updated to its latest release
(9575, 17 Feb. 2017).  The binary evolution code is used to
investigate a number of effects not present in the dynamical
simulations, in particular the effect of the tidal interaction on the
synchronization and circularization time.
We first created Zero Age Main Sequence (ZAMS) models starting from a
pre-main sequence model with a specific metallicity and helium abundance
($Z=0.01$, $Y=0.29$ and $Z=0.03$, $Y=0.26$, selected to match the two populations of Terzan 5, see Figure~\ref{fig:zam}). 

Then we evolved a set of 10 stars with mass 0.8, 0.9, 1.0, 1.1 and
1.2$\msun$ for about 11 and 5 Gyr for the low and high metallicity
population, respectively.  Then we used the evolved stellar models in
a grid of binaries with initial orbital period between 10 and 30 hours
and with an accretor mass between 1.2 and 2.3$\msun$. The binary
eccentricity was also included as a parameter, spanning values between
0 and 0.95.  Since we have found that the dynamical encounters will
perturb the orbit (along with the eccentricity) of \source on average
every 30 Myr, we want to explore which of our binaries lose their
eccentricity to within the currently observed upper limits of
$e<0.001$~\citep{pap12,pat12b}.  The initial stellar angular rotation
is treated as uniform and is taken to vary between
$\Omega_r=10^{-6}\rm\,rad\,s^{-1}$ and
$\Omega_r=6.1\times10^{-5}\rm\,rad\,s^{-1}$. A summary of the grid of
values used is reported in Table~\ref{tab:grid}.

\begin{table}
\begin{center}
\caption{Binary Evolution Grid.}
\leavevmode
\begin{tabular}{lccc}
\hline
Parameter                  & Initial Value & Final Value & Stepsize\\
\hline
Orbital Period~(hours)           &   15 & 31 & 2\\
Eccentricity               & 0 & 0.95 & 0.2\\
Donor Mass~($\msun$)                &  0.8 & 1.2 & 0.1\\
Neutron Star Mass~($\msun$)       & 1.2 & 2.1 & 0.15 \\
Donor Angular Velocity~($\rm\,rad\,s^{-1}$)  & $10^{-6}$ & $6.1\times10^{-5}$ & $10^{-5}$ \\
\noalign{\vspace{0.1cm}} 
\hline
\end{tabular}\label{tab:grid}
\end{center}
\footnotesize{Orbital and donor parameters used to evolve the $\approx10,000$ binaries used in this work. The last step-size of the eccentricity is taken to be 0.15 instead of 0.2. 
}
\end{table}
%%%%%%%%%%%%%%%%%%%%%%%%%%%%%%%%%%%%%%%%%%%%%%%%%

The circularization of the orbit and the synchronization of the
companion rotation occurs via tidal interaction and is treated
according to~\citet{hut81} for stars with a convective envelope.  The
tidal evolution of the orbit proceeds on a timescale:
\begin{equation}
  T = \frac{R^3}{G\,M\,\tau}=\frac{1}{4\pi^2}\left(\frac{P_s}{\tau}\right)\,P_s
\end{equation}
where $M$ and $R$ are the donor mass and radius, $G$ is the
gravitational constant, $\tau$ is a lag used in the weak friction model
(see \citealt{dar79, hut81}) and $P_s$ is the orbital period of a
particle grazing the surface of the donor star.

\section{Results}\label{sec:res}

As a consequence of dynamical encounters, the binaries in Terzan 5  can
undergo exchange interactions, during which one of the two stars of
the binary is exchanged with an incoming star. Dynamical processes can
also create new binaries when a star passes sufficiently close to
another star so that tidal interactions can bind the two objects
together.  Furthermore, even if the exchange or the tidal interactions do
not take place in large number, we need to consider the occurrence of
a large number of stellar fly-byes. Finally, if the accretor is a
massive white dwarf with stable hydrogen burning, then its mass can
increase above the Chandrasekhar limit and collapse to form a neutron star. 
In the following, we thus consider these three scenarios for the formation
of \source. 

\subsection{Scenario I: Exchange Interaction}\label{sec:exc}

%%%%%%%%%%%%%%%%%%%%%%%%%%%%%%% TABLE 1%%%%%%%%%%%%%%%%%%%%%%%%%%%%%%%%%
\begin{table}
\begin{center}
\caption{Statistics of exchanges.}
\leavevmode
\begin{tabular}{ll}
\hline
 $P_{\rm exch}$                        & 0.62 \\
 $P_{\rm exch,\,{}2}$                  & 0.59 \\
 $\langle{}m_{\rm 2,\,{}fin}\rangle{}$ [M$_\odot{}$] & 1.07 \\
 $\langle{}N_{\rm exch}\rangle{}$      & 1.22 \\
 $\tau{}_{\rm exch}$ [Gyr]             & 9.84 \\
\noalign{\vspace{0.1cm}}
\hline
\end{tabular}\label{tab:1}
\end{center}
\footnotesize{ $P_{\rm exch}$ is the probability that a simulated
  binary makes at least one exchange during the simulation ($t=12$
  Gyr). $P_{\rm exch,\,{}2}$ is the probability that a simulated
  binary makes at least one exchange in which the secondary star is
  ejected, whereas the neutron star remains in the system, during the
  simulation. $\langle{}m_{\rm 2,\,{}fin}\rangle{}$ is the average
  mass of the companion at the end of the simulation.
  $\langle{}N_{\rm exch}\rangle{}$ is the average number of exchanges
  per binary (including even binaries that do not undergo any
  exchanges), during the simulation.  $\tau{}_{\rm exch}$ is the
  exchange timescale.}
\end{table}
%%%%%%%%%%%%%%%%%%%%%%%%%%%%%%%%%%%%%%%%%%%%%%%%%%%%%%%%%%%%%%%%%%%%%%%%%%%%%

Table~\ref{tab:1} shows that exchanges are very frequent for the considered
binaries: about 62\% of the simulated binaries undergo at least one
exchange during the simulation. The outcome of an exchange is the ejection
of the secondary star in 95\% of cases (and the neutron star retained in the
system) which can be easily understood as the neutron star is by far the most
massive object in the initial binary.

The average mass of the secondary at the end of the simulation is
$\langle{}m_{\rm 2,\,{}fin}\rangle{}\simeq1.1$ M$_\odot{}$. We remind that
the initial average mass of the secondary (given the assumed Salpeter
initial mass function and mass range) is $\langle{}m_{\rm 2,\,{}fin}\rangle{}=0.23$
M$_\odot{}$. This implies a significant increase of the secondary mass
as a consequence of exchanges.

Finally, the average number of exchanges per binary during the entire
simulation is $\langle{}N_{\rm exch}\rangle{}=1.22$. This implies that
the exchange rate from the simulation is $\nu{}_{\rm
  exch}=1.02\times{}10^{-10}$ yr, i.e. that the exchange timescale is
$\tau{}_{\rm exch}\equiv{}\nu{}_{\rm exch}^{-1}\simeq 10$ Gyr.
%%%%%%%%%%%%%%%%%%%%%%%%%%%%%%%%%%% FIGURE
%%%%%%%%%%%%%%%%%%%%%%%%%%%%%%%%%%% 1 %%%%%%%%%%%%%%%%%%%%%%%%%%%%%%%%%%
\begin{figure}
\center{{
\epsfig{figure=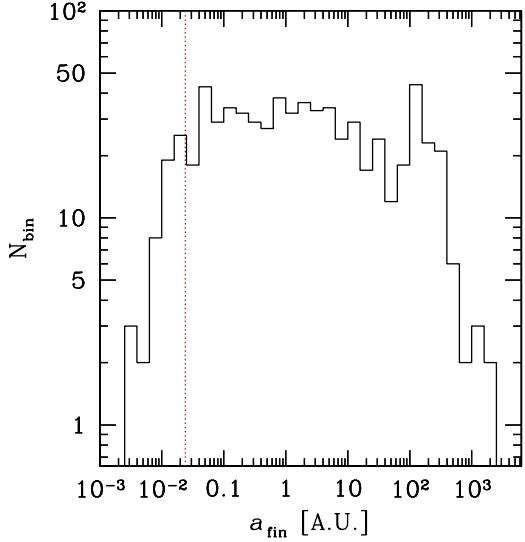,height=7.5cm}
}}
\caption{\label{fig:aorb} Distribution of semi-major axes at the end
  of the simulation. On the $y-$ axis, $N_{\rm bin}$ is the number of simulated binaries. The vertical dotted line indicates the current semi-major axis of \source.}
\end{figure}
%%%%%%%%%%%%%%%%%%%%%%%%%%%%%%%%%%%%%%%%%%%%%%%%%%%%%%%%%%%%%%%%%%%%%%%%%%%%%%%
%%%%%%%%%%%%%%%%%%%%%%%%%%%%%%%%%%% FIGURE 2 %%%%%%%%%%%%%%%%%%%%%%%%%%%%%%%%%%
\begin{figure}
\center{{
\epsfig{figure=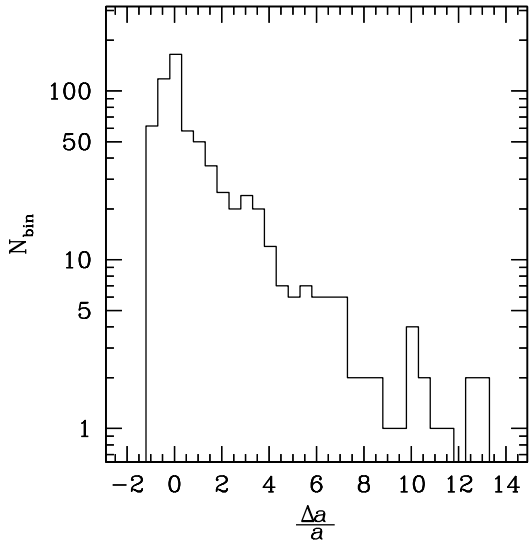,height=7.5cm}
}}
\caption{\label{fig:delta} Distribution of the relative changes of
  semi-major axis at the end of the simulation. On the $y-$ axis, $N_{\rm bin}$ is the number of simulated binaries.}
\end{figure}
%%%%%%%%%%%%%%%%%%%%%%%%%%%%%%%%%%%%%%%%%%%%%%%%%%%%%%%%%%%%%%%%%%%%%%%%%%%%%%%
%%%%%%%%%%%%%%%%%%%%%%%%%%%%%%%%%%% FIGURE 3 %%%%%%%%%%%%%%%%%%%%%%%%%%%%%%%%%%
\begin{figure}
\center{{
\epsfig{figure=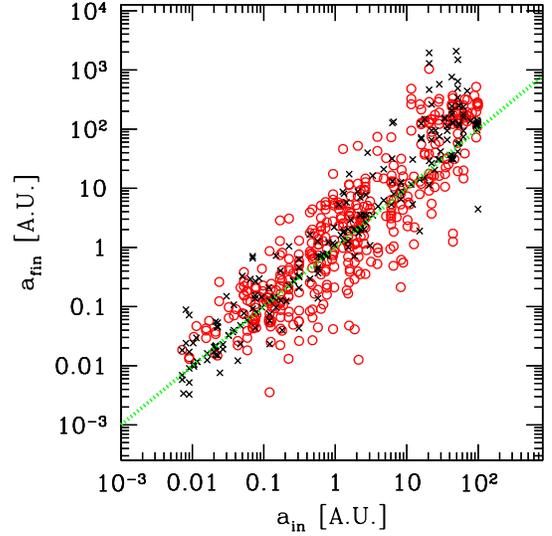,height=7.5cm}
}}
\caption{\label{fig:ainfin} Final versus initial semi-major
  axis. Crosses: simulated binaries that never underwent exchanges;
  open circles (red): simulated binaries that underwent
  exchange. Dotted line (green): line with $a_{\rm
    fin}=a_{\rm in}$.}
\end{figure}
%%%%%%%%%%%%%%%%%%%%%%%%%%%%%%%%%%%%%%%%%%%%%%%%%%%%%%%%%%%%%%%%%%%%%%%%%%%%%%%
%%%%%%%%%%%%%%%%%%%%%%%%%%%%%%%%%%% FIGURE 4 %%%%%%%%%%%%%%%%%%%%%%%%%%%%%%%%%%
\begin{figure}
\center{{
\epsfig{figure=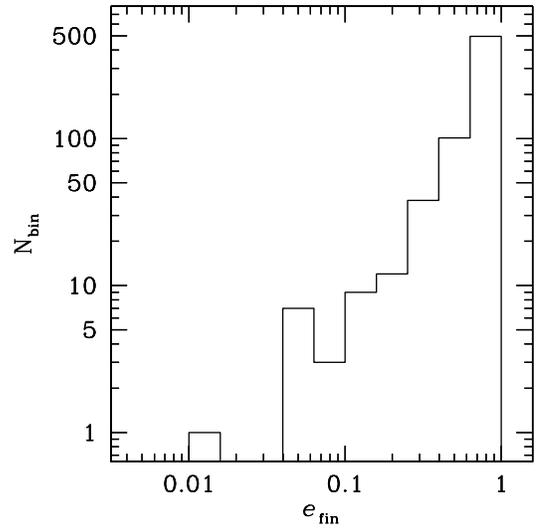,height=7.5cm}
}}
\caption{\label{fig:decc} Distribution of eccentricity at the end of
  the simulation. On the $y-$ axis, $N_{\rm bin}$ is the number of simulated binaries. }
\end{figure}
%%%%%%%%%%%%%%%%%%%%%%%%%%%%%%%%%%%%%%%%%%%%%%%%%%%%%%%%%%%%%%%%%%%%%%%%%%%%%%%

\subsubsection{Binary Orbital Properties}\label{sec:bin}

The simulations give us insight about the orbital properties of the
binaries, and in particular about the final semi-major axis $a_{\rm
  fin}$ and eccentricity $e_{\rm fin}$. The values of $a_{\rm fin}$
and $e_{\rm fin}$ that we show must be considered as values before the
start of mass transfer, as the BEV code does not include mass transfer
and circularization.  Figure~\ref{fig:aorb} shows the distribution of
$a_{\rm fin}$ (we remind that the distribution of $a_{\rm in}$ was
uniform in logarithm and in the $7\times{}10^{-3}$--$1\times{}10^{2}$ AU
range). The final distribution is still approximately uniform in
logarithm (except for the low/high $a_{\rm fin}$ tails) and spans the
$2\times{}10^{-3}-2\times{}10^{3}$ AU range. The vertical dotted line
in Figure~\ref{fig:aorb} shows the current orbital separation of
\source ($\approx0.024$ AU for an assumed total system mass of 2.3$\msun$), which is in the allowed range.
Figure~\ref{fig:delta} shows the relative change of the semi-major axis
$a$, defined as ${\rm d}a/a\equiv{}(a_{\rm fin}-a_{\rm in})/a_{\rm
  in}$: $\sim{}66$ per cent of binaries widen, as a consequence of the
interactions. $a_{\rm fin}$ versus $a_{\rm in}$ are shown in
Figure~\ref{fig:ainfin}, for binaries that underwent exchange (open
circles) and for binaries that did not undergo exchange
(crosses). There is no significant difference between systems that
underwent at least one exchange and systems that never underwent
exchanges. On the other hand, there are two interesting trends. First,
binaries that undergo exchanges can shrink more than binaries that do
not exchange their members. This is a consequence of the larger
binding-energy gain connected with exchanges. Second, it is unlikely
that very hard binaries (with $a_{\rm in}\lesssim{}0.01$ AU) undergo
exchanges.
The final distribution of eccentricity ($e_{\rm fin}$; Figure~\ref{fig:decc}) shows a
depletion of binaries at low ($<0.1$) eccentricity (initial
eccentricities were generated according to a thermal distribution
between 0 and 1). This is a consequence of the high frequency of
exchanges (which generally lead to an increase of eccentricity, see,
e.g., \citealt{sig93}), combined with the absence of circularization
recipes in the code. This fact is confirmed by looking at the relative
change of eccentricity, defined as ${\rm d}e/e\equiv{}(e_{\rm
  fin}-e_{\rm in})/e_{\rm in}$: about 72 per cent of the simulated
systems have significantly higher eccentricity by the end of the
simulation.

Another important result is that the relative variation of the
semi-major axis \textit{per encounter} is $ \left|\frac{\Delta
    a}{a}\right| \simeq 0.15$ so that on average, the semi-major axis
changes by about $15\%$ for each encounter of the binary.  Since each
binary has an average of one encounter every 32 Myr, the total number
of encounters each binary has done during the entire lifetime of the
cluster is $N_{enc}\simeq 400$.  The binary semi-major axis has been
perturbed enormously during the lifetime of the system, with the
companion star being pushed closer or further away from the compact
object at a high rate. Despite this huge variation of the orbital
separation per encounter, the final orbital separation after $\sim375$
encounters is still rather close to the initial orbital separation
because the ratio between hardening and softening encounters is close
to unity.

The binaries which have $a_{fin}<a_{in}$ change their semi-major axis,
on average, by a fraction: $\left\langle{}\frac{\Delta
  a}{a}\right\rangle{} \simeq -0.5$ whereas those with
$a_{fin}>a_{in}$ drift by a total amount of
$\left\langle{}\frac{\Delta a}{a}\right\rangle{} \simeq 3$.  This
means that in the cluster lifetime ``hard'' binaries have a final
orbital separation which is about 50\% of their initial value, whereas
``soft'' binaries have a final orbital separation 4 times larger than
the initial value.

\subsubsection{Binary Evolution}\label{sec:bev}

If \source has been formed via an exchange interaction, then the
binary needs to still dissipate its eccentricity (on average) within
the next close encounter. To verify whether this is possible, we have
looked at our binary evolution calculations.
We find
that none of our binaries is able to circularize within the 30 Myr
encounter timescale, unless the
initial binary eccentricity is already smaller than
$\lesssim0.2$. This means that the exchange interaction model requires
substantial fine-tuning of the initial parameters.

\subsection{Scenario II: Tidal Capture}\label{sec:tid}

Tidal interactions are not included in our dynamical calculations but
they can nonetheless provide a valid scenario to explain the
properties of \source.  A neutron star can capture an incoming star by
deforming the stellar structure via tides that remove part of the
kinetic energy of the incoming star. This mechanism might in principle
form moderately tight binaries with orbital separations of 3--4 tidal
radii \citep{mcm87,mar96}.  This can be defined as:
\begin{equation}
r_{t} = \left(\frac{m_1}{m_2}\right)^{1/3}R_{2}
\end{equation}
where $R_2$ and $m_2$ are the radius and mass of the incoming star and
$m_{1}$ is the neutron star mass.  If we assume a canonical 1.4
$\msun$ neutron star mass and we use the actual observed and inferred
donor parameters -- $m_2\simeq1$, $R_{2}\simeq1.3\,R_{\odot}$ (see for
example~\citealt{pat12b}), then a binary with an orbital separation of
0.014--0.03 AU can be formed via tidal capture, which is within 
the range of the currently observed value.

\begin{figure}
\centering
\rotatebox{0}{\includegraphics[width=1.0\columnwidth]{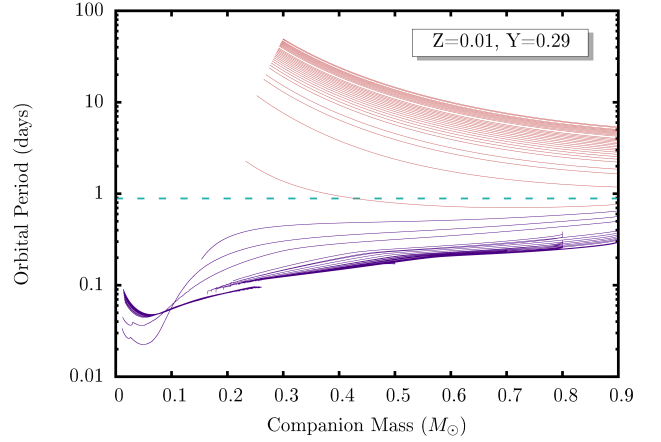}}
\caption{Bifurcation period for $\approx 100$ models with a 1.4$\msun$ neutron
  star and a companion with mass 0.5--0.9$\msun$. The red lines are diverging systems whereas the blue ones are converging ones. The dashed horizontal line identifies the orbital period of \source ($\approx0.8875$ d), which should thus be a diverging system and is very close to the bifurcation period of about 0.7--0.8 days.}\label{fig:bif}
\end{figure}

Support to the tidal capture scenario comes from the fact that the
orbital period of \source is surprisingly close to the bifurcation
period (see Figure~\ref{fig:bif}).  Indeed all neutron star binaries created via
tidal capture are formed with a period close to the bifurcation
period, provided that their mass and radius are close to solar
values~\citep{pod02,dis92}. The tidal capture scenario provides
therefore a natural explanation to the orbital parameters of \source,
but it remains to be verified whether the capture rate is sufficiently
high to guarantee a high probability for such an event.

Further support for the tidal capture scenario comes from the low eccentricity
of the system. Indeed if \source is formed via tidal capture then it
will be a very hard binary, meaning that its orbital energy is much
larger than the average kinetic energy of the incoming stars interacting in close
encounters. This guarantees a very quick circularization of the orbit.
The current orbital binding energy of \source is:
\begin{equation}
E_b = \frac{G\,m_1\,m_2}{2\,a}
\end{equation}
which, for $m_1=1.4\msun$ and $m_2=1\msun$ gives a value of the order of $5\times10^{47}\rm\,erg$. The average kinetic energy of a star in the cluster is:
\begin{equation}
E_k=\frac{1}{2}\langle m\rangle\sigma_c^2{\sim}10^{45}\rm\,erg
  \end{equation}
where $\langle m\rangle$ is the average stellar mass in the
cluster. Since $E_k\ll~E_b$, \source is indeed a hard binary.

\begin{figure}
\centering
\rotatebox{0}{\includegraphics[width=1.0\columnwidth]{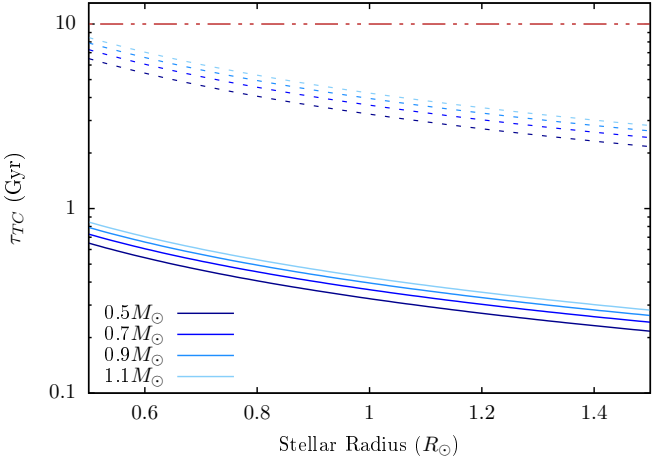}}
\caption{Tidal capture timescale for a 1.4$\msun$ neutron star and four different donors with radii spanning a range between 0.5 and 2$R_{\odot}$.
The solid lines refer to a neutron star globular cluster core density of $n_{NS}=10^{3}\rm\,pc^{-3}$, whereas the dashed lines to $n_{NS}=10^{2}\rm\,pc^{-3}$. 
The horizontal dot-dashed red line marks the exchange timescale $\tau_{\rm exch}\approx10$ Gyr. The tidal capture timescale is substantially shorter than the exchange interaction timescale in all cases.}\label{fig:tid}
\end{figure}

The tidal capture rate can be estimated as:
\begin{equation}
\Gamma_{TC}\simeq\frac{4\pi\,r_c^3}{3}n_{NS}n_{c}\Sigma\,\sigma_c
\end{equation}
where $r_c$ is the core radius of the cluster (0.16 pc in our case), $n_{NS}$ is the number densities of neutron stars in the cluster core, $n_c$ is the cluster core density (see Section~\ref{sec:dyn}) and $\Sigma$ is the cross section of the interaction. This latter quantity is defined as:
\begin{equation}
\Sigma = \pi\,r_t^2\left(1+\frac{2G\,m_1}{\sigma_c^2\,r_t}\right).
\end{equation}
The largest uncertainties in the calculation of this expression come
from the poor knowledge of the total number of neutron stars in the
core of Terzan 5.  However, since we know that the cluster contains
several tens of pulsars and X-ray binaries and it has even been
suggested it might contain few hundred pulsars~\citep{abr11}, we can
use a lower limit of the order of $n_{\rm NS}{\sim}10^{2}$--$10^3$ pc$^{-3}$
to estimate $\Gamma_{TC}$. This gives a tidal capture timescale of
$\tau_{TC}{\sim}0.2$--8 Gyr, depending on the stellar mass and
radius (see Figure~\ref{fig:tid}).
If we consider also stars that have evolved significantly off the
main sequence, then the tidal capture timescale can be as short as a
few tens of million years (for a $R=10~R_{\odot}$ giant star).
In this case the companion of \source might be a stripped giant star,
something proposed already for other sources in globular clusters (e.g., \citealt{zyl04}).
However, even when we consider main sequence and sub-giant stars with
a mass and radius compatible with those of the donor in \source we
find that $\tau_{\rm \, TC}$ is much shorter of both the age of the
cluster and the exchange interaction timescale $\tau_{exc}$ calculated
in the previous sections.

Finally, the eccentricity of binaries formed via tidal capture is known to
dissipate on a timescale which is much shorter than any other
timescale considered in this work~\citep{mcm87}, even when considering
the possible initial chaotic state of the orbits~\citep{mar95a,
  mar95b}. This again favors the tidal capture scenario because it can
easily explain the basically circular orbit observed.

\subsection{Scenario III: Accretion Induced Collapse}\label{sec:aic}

Accretion induced collapse (AIC) has been invoked to solve some
problems related with the fine tuning of the short lived wind fed and
RLOF phase of \source \citep{pat12b}.  In our simulations, however, we
have demonstrated that the dynamical encounters are in principle
sufficient to explain the observations, with the exception of the
rather low magnetic dipole moment of the neutron star, under the
assumption that the field decays via fluxoid expulsion from the
neutron star core during the wind fed epoch.

The variation of the orbit generated by a quasi-spherical mass loss is
given by: \be \frac{a_{fin}-a_{in}}{a_{in}}=\frac{\Delta M}{M_1 + M_2 - 2 \Delta
  M} \ee where $a_i$ and $a_f$ are the initial and final orbital
separation once the collapse has occurred, $\Delta M\sim 0.2\msun$ is
the amount of mass lost in the collapse and $M_1 + M_2$ is the total
mass of the binary prior the collapse.  Since $M_1$ is constrained to
be the Chandrasekhar mass of the white dwarf and $M_2$ cannot be much
different than $\sim1\msun$ since the star belongs to the globular
cluster, the expansion of the orbit would be of the order of
$10-20\%$. This value (similar to the variation of the semi-major axis
produced by a single encounter, see Section~\ref{sec:bin}) has
therefore little effect on the evolution of the binary.

The AIC scenario requires that the formation of the neutron star has
happened not too early in the lifetime of the globular cluster, since
otherwise the evolution of the binary would still be dominated by the
dynamics in the globular cluster, and invoking the AIC scenario would
thus be unnecessary.  The mass transfer phase should therefore be
relatively short and the initial white dwarf mass needs to be rather
massive ($\gtrsim 1\msun$) so that the donor star has enough material
to bring the white dwarf above the Chandrasekhar limit. Furthermore
the donor has to still preserve a mass of about $0.8-0.9\msun$ to
feed the neutron star in the currently observed RLOF epoch. The initial mass
of the donor is thus required to be at least $1.0-1.3\msun$ and
 the AIC must have happened when these kind of stars were
 still present in the cluster.
 
Furthermore, for the AIC to occur, the mass accretion rate from the donor star
towards the white dwarf needs to be within a certain range such that
hydrogen burning on the white dwarf is stable.  Explosive
ignition of the accreted matter would otherwise hardly allow any 
mass growth of the compact object since most of the mass would be
expelled during the explosion (\citealt{tow05},~\citealt{van11}). The
critical range of stable accretion weakly increases with the white
dwarf mass, and reaches values of the order of
$1-4\times10^{-7}\rm\,\msun\,yr^{-1}$ for a massive white dwarf of
1--1.2$\msun$. To have such large mass accretion rates a relatively massive
donor is required~(see e.g., \citealt{tau13}) which again would
conflict with the constraint that the binary has formed recently.
These particular set of requirements imply that the formation scenario
with an accretion induced collapse white dwarf is the most unlikely
among the three explored in this work. 

\section{Roche Lobe Overflow and the Effect of Close Encounters}\label{sec:rlo}

We now switch to the question of what happens to interacting binaries when a
dynamical encounter occurs after a contact phase has already started.
The effect of close encounters will indeed alter the orbital evolution
of the binary and produce sudden shifts in location of the donor
star. In essence, this can suddenly stop or trigger a RLOF phase.

Once the donor star has drifted sufficiently close to the neutron
star, a Roche lobe overflow (RLOF) phase can start.  Evolutionary
sequences of LMXBs have been extensively explored in the literature
(see for example \citealt{rap82, pod02}). However, the evolutionary
sequences of binaries perturbed by close encounters every $\sim30$ Myr
are likely to have a different outcome.  Close encounters recur on a
timescale which is much shorter than both typical nuclear and magnetic
braking evolutionary timescales of low mass main-sequence and giant
stars. The same holds if we compare with the timescale of
gravitational radiation emission, which is one to two orders of
magnitude longer. Furthermore, as we found in Section~\ref{sec:res},
the change in semi-major axis experienced at each encounter is of the
order of $15\%$ and it can be sufficient to detach the donor from its
Roche lobe and switch off the accretion, or bring the donor star into
contact (depending on the sign of the semi-major axis change) at an
earlier stage than the typical evolutionary timescale. The Roche lobe
size scales linearly with the orbital separation $R_L\propto a$ and
therefore each encounter changes the Roche lobe by an amount
proportional to the variation of the semi-major axis. It is also
possible that the donor is already in RLOF before the next encounter
occurs. The typical duration of non-resonating encounters lasts in
general less than a few orbital periods, so that the variation of the
Roche lobe size can be considered almost instantaneous. In this case
the mass transfer rate is suddenly increased or halted, depending on
the sign of $\dot{a}$.

If the star is already in (or close to) a RLOF phase, then
a close encounter that shrinks the orbit would move the star off hydrostatic equilibrium in a very short time. Indeed the slope of the mass-radius relation for the Roche lobe $\zeta_L$ would be much larger than the slope of the same relation for the donor star ($\zeta_d$):
\begin{equation}
\zeta_L = \frac{d{\rm\,ln\,R_L}}{d{\rm\,ln\,m_2}} \gg \zeta_d = \frac{d{\rm\,ln\,R}}{d{\rm\,ln\,m_2}}.
\end{equation}
Therefore we do not follow this effect in detail since dynamical timescales need to be considered.  We can anyway
speculate that four scenarios are likely to appear:
\begin{itemize}
\item the encounter shrinks/widens the orbit of a detached system which
remains detached
\item the encounter widens the orbit of a contact systems and shuts down
the mass transfer phase
\item the encounter shrinks the orbit of a detached system and brings
the donor into Roche lobe contact
\item the encounter shrinks the orbit of a contact system and produces
an enhancement of the mass transfer rate
\end{itemize}

The latter two possibilities are perhaps the most interesting as they
might produce an abrupt increase of the mass transfer rate well in
excess of the Eddington limit possibly leading to a common envelope
(CE) phase or an ejection of the stellar envelope. If we assume that a
certain fraction of envelope mass is lost in the encounter, then the
orbital energy variation per encounter is of the order
of~\citep{iva11}:
\begin{equation}
\Delta\,E_{orb} = -\frac{G\,m_1\,m_2}{2\,a_{in}}+\frac{G\,m_1\,m_{2,c}}{2\,a_{fi}}  
\end{equation}
where $m_{2,c}$ is the donor mass after it has expelled (a fraction of) its envelope.
The typical binding energy of the donor envelope is instead:
\begin{equation}
  E_{bin} = -\frac{G\,m_2\,m_{2,e}}{R}
\end{equation}
By taking as a typical orbital separation a value of 0.01--0.02 AU, a
donor mass of $1\msun$ and an envelope mass of 0.1--0.5$\msun$, then
$\Delta\,E_{orb}$ and $E_{bin}$ are of the same order of magnitude (${\sim}10^{47-48}\rm\,erg$), meaning
that an encounter can lead to the ejection of the entire stellar envelope.

The details of the unstable mass transfer leading to a common envelope
phase is subject to more uncertainties, with the star that is
expected to be suddenly brought out of hydrostatic equilibrium. A
spiral-in phase can also set in, with the compact object merging with
the stellar core (see \citealt{pod01} for a discussion) possibly
producing Thorne-Zytkow type objects~\citep{tho77}.

Even if we do not follow such evolutionary phases in our calculations,
it is important to highlight that such unstable binaries are expected
to occur in large numbers \textit{because of the high rate of
  encounters} and therefore might be common in massive and
concentrated globular clusters like Terzan 5.  The question on whether
this process might be connected with the over-abundance of pulsars and
X-ray binaries in concentrated and massive clusters cannot be
addressed at this moment.

\section{Conclusions}\label{sec:con}

In this paper we have proposed a formation scenario for the peculiar
accreting binary \source in the globular cluster Terzan 5.  Our
calculations show that the most likely scenario for the formation of
\source is one involving a tidal capture of a star, which occurs on a
timescale of 0.2--8 Gyr (depending on physical details of the donor
star).  This scenario can explain naturally the apparently young age
of \source, the currently orbital separation of the binary, its
(nearly) circular orbit and the proximity of the orbital period to the
bifurcation one. This scenario requires no fine tuning, differently
from the other mechanisms considered in this work.  This conclusion is
also compatible with the earlier suggestion mae by \citet{pat12b} that
\source is not a primordial binary, since that scenario would be
incompatible with the spin-up behavior of the system and would require
fine tuning for the beginning of the RLOF phase.

Dynamical simulations suggest that \source, as well as other binaries
in the core of Terzan 5, have been strongly and constantly perturbed
by close encounters occurring on average every ${\approx}30$ Myr,
which might have altered their normal evolution.  These perturbations
might induce ``hiccups'' in the accretion phase, with the RLOF
suddenly switching on and off. This can happen only in dense
environments like globular cluster cores. It remains to be assessed
whether our evolutionary scenario correctly represents the observed
X-ray source populations in globular cluster.

\acknowledgments{AP acknowledges support from an NWO Vidi fellowship.
  MM acknowledges financial support from the Italian Ministry of
  Education, University and Research (MIUR) through grant FIRB 2012
  RBFR12PM1F, from INAF through grant PRIN-2014-14, and from the MERAC
  Foundation.  MM thanks S. Sigurdsson for providing the original
  version of the BEV code. AP thanks T.~Tauris, L.~Yungelson and A.~Milone for
  useful discussions and P.~Podsiadlowski for pointing out the
  importance of tidal interactions.}

%\appendix

\end{document}